\documentclass[fleqn,usenatbib]{mnras}

\usepackage[T1]{fontenc}
\usepackage{ae,aecompl}
\usepackage[dvipsnames]{xcolor}

\usepackage{graphicx}	
\usepackage{amsmath}	
\usepackage{amssymb}	
\usepackage{listings}
\usepackage{soul,xcolor}

\title[Early halos and PBH merger rate]{Influence of early dark matter halos on the primordial black holes merger rate}

\author[V. Stasenko and K. Belotsky]{Viktor Stasenko$^{1,2}$\thanks{vdstasenko@mephi.ru} and Konstantin Belotsky$^{1,2}$ \thanks{kmbelotskij@mephi.ru}\\
\\
$^{1}$National Research Nuclear University ``MEPhI'', 115409, Moscow, Russia\\
$^{2}$Novosibirsk State University, 630090, Novosibirsk, Russia
}

\date{Accepted XXX. Received YYY; in original form ZZZ}

\pubyear{2023}

\begin{document}
\label{firstpage}
\pagerange{\pageref{firstpage}--\pageref{lastpage}}
\maketitle

\begin{abstract}
Primordial black hole (PBH) binaries forming in the early Universe may contribute to the merger events observed by the LIGO-Virgo-KAGRA collaborations. Moreover, the inferred merger rate constraints the fraction of PBH with masses $m \sim 10 \, M_{\odot}$ in the dark matter (DM) to $f_{PBH} \lesssim 10^{-3}$. This constraint assumes that after the formation of PBH binaries, they do not get destroyed or their parameters are not perturbed until the merger. However, PBHs themselves contribute to the formation of early DM structures in which the interactions between PBHs take place actively. This leads to the fact that the binaries can be perturbed in such a way that their lifetime becomes longer than the Hubble time $t_H$. In this work, we consider the effect of the initial spatial Poisson distribution of PBHs on the structure formation at the high redshifts $z \gtrsim 10$. Next, we explore the evolution of such halos due to the interaction of PBHs with each other and with DM particles. We show that the early halos evolve on timescales much shorter than the age of the Universe. Furthermore, for fractions of PBHs $f_{PBH} < 1$, the internal dynamics of a halo is significantly accelerated due to the dynamical friction of PBHs against DM particles. As a result, a significant fraction of binaries will be perturbed in such structures, and the gravitational waves constraints on PBHs with masses $m \sim 10 \, M_{\odot}$ can be weakened to $f_{PBH} \sim 0.1$. 

\end{abstract}



\section{Introduction}

Merging primordial black holes (PBHs) binaries are actively considered as sources of gravitational wave events observed by the LIGO-Virgo-KAGRA collaborations \citep{Sasaki:2016jop, Bird:2016dcv, Clesse:2016vqa, Blinnikov:2016bxu, Raidal:2017mfl, Ali-Haimoud:2017rtz, Sasaki:2018dmp, Raidal:2018bbj, Hutsi:2020sol, Dolgov:2020xzo}. However, in addition to gravitational waves, PBHs can explain a number of other modern problems of cosmology and astrophysics \citep{Clesse:2017bsw, Carr:2021bzv, 2023arXiv230603903C}. Among them are the observation of supermassive black holes (quasars) at high redshifts $z \gtrsim 6$ \citep{Banados:2017unc, Inayoshi:2019fun, 2020ApJ...897L..14Y, 2021ApJ...907L...1W}, possibility of explaining some part or even all of the dark matter (DM) depending on the PBH mass \citep{Green:2020jor, Carr:2021bzv, 2023arXiv230603903C}. Galaxies discovered by the JWST telescope in the very young Universe \citep{2022ApJ...938L..15C, Ferrara:2022dqw, 2022ApJ...940L..55F, 2023MNRAS.519.1201A, 2023Natur.616..266L} can be also explained in terms of PBH \citep{Liu:2022bvr, Hutsi:2022fzw, 2023arXiv230101365D}. 

Despite the fact that PBHs can potentially be part of gravitational wave events, the observed merger rate imposes the strongest constraints on the fraction of PBHs in DM $f_{PBH} \lesssim 10^{-3}$ at tens of solar masses \citep{Sasaki:2016jop, Ali-Haimoud:2017rtz, Kavanagh:2018ggo, Pilipenko:2022emp, Postnov:2023ntu, Jangra:2023mqp}. This constraint suggests that in the early Universe, a pair of PBHs decouple from the Hubble flow and form a binary system \citep{Nakamura:1997sm, Ioka:1998nz}. Further, these binaries gradually undergo the stage of inspiral due to the emission of gravitational waves and eventually merge. However, from the moment of formation to the merger, the parameters of the binary system can be significantly perturbed so that the lifetime of the binary exceeds the Hubble time $t_H$, as a result of which the merger rate can be reduced \citep{Vaskonen:2019jpv, Jedamzik:2020ypm, DeLuca:2020jug}. In addition, if PBHs are initially strongly clustered, the constraint on gravitational waves can be relaxed to $f_{PBH} \sim 1$ \citep{Eroshenko:2023bbe}. 

The discrete nature of PBHs leads to the early formation of structures due to their initial Poisson distribution in space \citep{1975A&A....38....5M, Afshordi:2003zb, Inman:2019wvr}. The forming halos have a much larger mass than predicted in purely adiabatic inflationary fluctuations with nearly scale invariant power spectrum. The evolution of such DM halos containing some amount of PBHs is, in many respects, similar to that of globular star cluster; in particular, for very early halos, the phenomenon of core collapse takes place (see for review \citet{1987degc.book.....S}), 
time of which is much less than the age of the Universe $t_H$. If a PBH binary finds itself in a halo that experiences core collapse, 
then it is 
very likely that it will not contribute to the PBH merger rate \citep{Vaskonen:2019jpv}. This is due to the fact that in the process of halo evolution, the probability of perturbing a pair of PBHs increases significantly. However, the analysis of \citet{Vaskonen:2019jpv} is somewhat simplified, because it does not take into account the halo density profile and an influence of dark matter particles on the dynamics of PBHs in the halo for the case $f_{PBH} < 1$. PBHs will experience dynamical friction against DM particles, which leads to a decrease in the core collapse time $t_{cc}$. The main aim of this article is to find the fraction of binaries that will be in halos that experience core collapse in a time less than the age of the Universe. 
In this paper, to study the listed effects, the kinetic Fokker-Planck equation is solved. This equation describes the evolution of a self-gravitating system and is often used to study the dynamics of globular star clusters and galactic nuclei \citep{2017ApJ...848...10V}. 

In this work, we generally consider the two-component dark matter, which consists of a PBH with mass \mbox{$m = 10 \, M_{\odot}$} (we assume a monochromatic mass spectrum) and some massive unknown particles that gravitationally interact with the PBHs. First, we show how PBHs influence the formation of early DM halos at high redshifts ($z \gtrsim 10$) due to Poisson noise. Then, the evolution of such halos is considered by solving the Fokker-Planck equation. We show that the timescale of the core collapse depends significantly on the PBH fraction in the DM composition $f_{PBH}$. Next, we qualitatively consider the dynamics of PBH binaries that form in the early Universe and show that their parameters can be significantly perturbed during halo evolution. Eventually, we show how the modern PBH merger rate is changing and conclude that the constraints can be relaxed to $f_{PBH} \sim 0.1$. 

\section{Early dark matter halos}

The discrete nature of PBHs induces Poisson fluctuations with amplitude $\delta \sim f_{PBH} / \sqrt{N}$, where $N$ is the average number of PBHs in the considered volume \citep{Carr:2018rid}. Note that these fluctuations are isocurvature perturbations and can have a much larger amplitude than the adiabatic inflationary fluctuations $\delta_{ad} \sim 10^{-5} \div 10^{-4}$. As a result, on small scales, the matter power spectrum is modified as 
\begin{equation}
    P(k) = P_{PBH} + T^2_{ad}(k) P_{ad}(k).
\end{equation}
Here $P_{ad} = A k^n$ with $n = 0.96$\footnote{Throughout this article, we use the standard cosmological $\Lambda$CDM model with parameters $\Omega_{CDM} = 0.27$, $\Omega_B = 0.05$, $\Omega_{\Lambda} = 0.68$.} \citep{Planck:2018vyg} and $T_{ad}$ are the matter power spectrum and the transfer function for adiabatic inflationary fluctuations \citep{Mo:2002ft} and $P_{PBH}$ is the PBHs contribution \citep{Mena:2019nhm, Hutsi:2019hlw}:
\begin{equation}
    P_{PBH} = \frac{f^2_{PBH}}{n_{PBH}} T^2_{iso},
\end{equation}
where $n_{PBH}$ is the comoving PBH number density and $T_{iso}$ is the transfer function for isocurvature perturbations, which describes the linear growth of fluctuations \citep{1999coph.book.....P, 2011itec.book.....G} 
\begin{equation}
    T_{iso} = \frac{3}{2} (1 + z_{eq}) g(0)
\end{equation}
where $g(0) \approx 0.74$ reflects the suppression of density fluctuation growth due to the $\Lambda$ term \citep{Mo:2002ft} and $z_{eq} \approx 3402$ is the redshift of matter-radiation equality. Thus, the power spectrum induced by the Poisson noise of PBHs has the form 
\begin{equation}
    P_{PBH} = \frac{9}{4} g^2(0) (1 + z_{eq})^2 \frac{m f_{PBH}}{\Omega_{DM} \rho_{crit}},
\end{equation}
where $\rho_{crit} \approx 126 \, M_{\odot}$~kpc$^{-3}$ is the critical density. 

The variance of fluctuations on the mass scale $M =  4 \pi R^3 \rho_{M} / 3$ as the function of the redshift $z$ is given by standard expression
\begin{equation} \label{sigma2_1}
    \sigma^2_M(R,z) = D^2(z)\int \frac{dk}{2 \pi^2} \, k^2 P(k) W^2 (kR),
\end{equation}
where function $D(z) = g(z) / [g(0) (1 + z)]$ describes the growth of fluctuations and $W(kR)$ is the window function in the Fourier space (Top-hat spherical filter)
\begin{equation}
    W(kR) = \frac{3 \, (\sin kR - kR \cos kR)}{(kR)^3}.
\end{equation}

On the scales where the main contribution to the power spectrum is due to PBHs (particularly at high redshifts $z$), the variance of fluctuations will be 
\begin{equation}
    \sigma^2_M(M,z) = \frac{9 \, m f_{PBH}}{4 \, M}  \left ( \frac{1 + z_{eq}}{1 + z} \right)^{2}.
\end{equation}
Here, for simplicity, the contribution of baryons is neglected, i.e.\ it is assumed that $\Omega_{M} = \Omega_{DM}$ and 
the value of the integral from the square window function
\begin{equation}
    \int_{0}^{\infty} dx \, x^2 W^2(x) = \frac{3}{2} \pi
\end{equation}
is used.

\begin{figure}
	\begin{center}
\includegraphics[angle=0,width=0.5\textwidth]{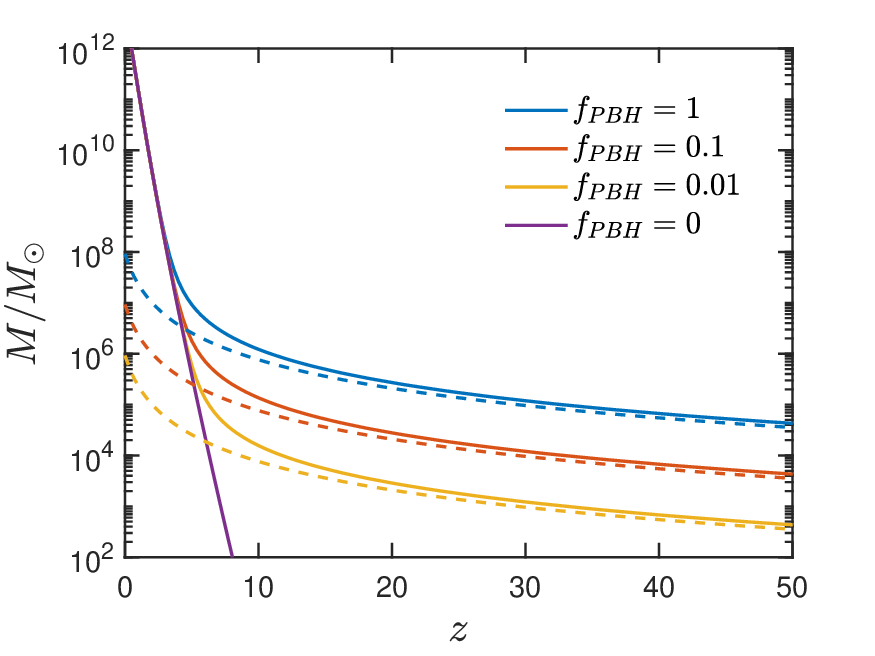}
	\end{center}
\caption{The characteristic halo mass as a function of the redshift for different fractions of PBHs. The solid lines correspond to the solution of the equation $\sigma_M(M,z) = \delta_c$, where $\sigma_M$ is defined by Eq.~\eqref{sigma2_1}. Dashed lines are given by the approximate Eq.~\eqref{M_char}.}
	\label{gr_Mh} 
\end{figure}

Let us define the characteristic halo mass (such halos are also called $1 \sigma$ halos) as $\sigma_M(M^*) = \delta_c$, where \mbox{$\delta_c = 1.69$} is the critical fluctuation value for spherical collapse calculated in linear theory. Fig.~\eqref{gr_Mh} shows the redshift dependencies of the characteristic halo mass calculated for different PBH contributions to DM $f_{PBH}$, where the dotted lines are obtained with the formula
\begin{equation} \label{M_char}
    M^* = \frac{9 \, m f_{PBH} }{4 \, \delta_c^2} \left ( \frac{1 + z_{eq}}{1 + z} \right)^{2},
\end{equation}
which approximates well at the redshifts $z \gtrsim 10$. It can be seen that the presence of PBHs leads to the production of more massive dark halos at high redshifts as compared with the case of purely adiabatic inflationary perturbations. 

In the spherical top hat collapse model, after the formation of a structure with mass $M$, the average density of a halo will be $\rho_H = \Delta_c \, \rho_{M}(z)$, where $\Delta_c = 18 \pi^2$
\begin{equation}
    \rho_H = \frac{3 M}{4 \pi R_{vir}^3} = 18 \pi^2 \rho_{crit} \Omega_M (1+z)^{3},
\end{equation}
where $R_{vir}$ is the virial radius. At high redshifts $z \gtrsim 10$ sufficiently dense halos are formed with a small velocity dispersion $\sigma \sim 1$~km/s, which leads to active interaction of PBHs with each other. The dynamical evolution of such halos is similar to that of globular star clusters, in particular, the core collapse occurs for them \citep{1968MNRAS.138..495L, 1980ApJ...242..765C} that is shrinking of the central region with increasing in density, it further increases the rate of PBH interactions. However, the structures are formed hierarchically from the bottom up, which can lead to the destruction of small halos in the process of their absorption by large structures. It is a difficult task to quantitatively take into account the dynamic internal evolution of halos and their interaction with each other during the structure formations. In this paper, we accept the common assumption that the dark halos are not destroyed with time and focus only on the internal dynamics of the halo and its influence on the PBH merger rate.

\section{Dynamics of PBH in the early structures}

Two processes affect the dynamics of PBHs in a halo: the interaction of PBHs with each other and with DM particles. Both these processes can be characterized by diffusion coefficients using the Fokker-Planck equation \citep{2008gady.book.....B}. They show the average rate of change in the body's velocity (PBH in our case) as a result of many weak gravitational encounters with other objects. The interaction between PBHs is characterized by the following coefficient
\begin{equation}
    \langle (\Delta v)^2 \rangle = \frac{4 \sqrt{2} \pi G^2 \rho_{BH} m \ln{\Lambda}}{\sigma},
\end{equation}
where $\sigma$ is the one-dimensional velocity dispersion, $\rho_{BH} = \rho_H f_{PBH}$ is the density of PBHs in the halo center and $\ln \Lambda \sim 10$ is the Coulomb logarithm. The characteristic timescale of PBH interactions is called the relaxation time $t_r$, which is defined as $t_r = \sigma^2 / \langle (\Delta v)^2 \rangle$ (see details in \citet{2008gady.book.....B}) 
\begin{equation}
    t_r = \frac{0.34 \, \sigma^3}{G^2 m \rho_{BH} \ln{\Lambda}}.
    \label{trelax}
\end{equation}

If the halo consisted only of PBHs, then its evolution is not different from the dynamics of the simplest globular star cluster and proceeds according to the scenario of a gravitational catastrophe \citep{1968MNRAS.138..495L}: under the influence of pairwise interactions of PBHs, the central region of the halo (hereinafter core) is compressed to a sufficiently large density. The core collapse time is a certain number of relaxation times $t_{cc} = \beta t_{r}$, where the proportionality constant $\beta$ is the number of those times which depends on the choice of the initial density profile \citep{Quinlan:1996bw}. Evolution after collapse is driven by three-body interactions with the formation of binary systems, which stop the collapse of the core and lead to the self-similar expansion of the cluster \citep{1987ApJ...319..801L, Takahashi:1996bn, 2008gady.book.....B}. The merger rate of such binaries was estimated in \citet{Franciolini:2022ewd}, but this channel is less efficient than mergers of binaries formed in the early Universe. 

On the other hand, if a halo contains dark matter particles in addition to PBHs, then the interaction of PBHs with DM particles can be characterized by the dynamical friction coefficient \citep{2013degn.book.....M}
\begin{equation}
    \langle \Delta v \rangle = - \frac{4 \pi G^2 \rho_{DM} m \ln{\Lambda}}{\sigma^2},
\end{equation}
where $\rho_{DM} = (1 - f_{PBH}) \rho_H$ is the density of DM particles. Dynamical friction leads to the energy loss of PBHs, as a result of which they settle in the center of the halo. Similarly to the case considered above, the characteristic dynamical friction time can be defined as
\begin{equation} 
    t_{df} = \frac{3}{8} \sqrt{\frac{2}{\pi}} \frac{\sigma^3}{G^2 m \rho_{DM} \ln{\Lambda}}.
    \label{t_df}
\end{equation}

It is physically obvious that, for a small fraction of PBHs in the DM, their dynamics is determined only by the interaction with DM particles, and the scattering of PBHs on each other is negligible. In the intermediate case, the picture is as follows: in the early stages of evolution, dynamical friction is the dominant process, as a result of which PBHs will sink into the center of the halo. When a sufficiently high concentration of PBHs is reached, their further dynamics will already be determined by pair interactions with each other. From the condition that the diffusion coefficients are equal $\sigma \langle \Delta v \rangle \sim \langle (\Delta v)^2 \rangle$, one can find the density of PBHs, starting from which the dynamics will be determined only by the pairwise interaction of PBHs. It can be seen that this occurs at $\rho_{BH} \sim \rho_{DM}$, which was obvious from physical considerations. PBH settling occurs on the scale of dynamical friction time Eq.~\eqref{t_df}. Further, as it was outlined, the evolution will occur due to interactions of PBHs with each other on the relaxation timescale Eq.~\eqref{trelax}, where $\rho_{BH} \approx \rho_{DM}$. Therefore it is convenient to choose the following quantity
\begin{equation} 
    t_{ch} = \frac{\sigma^3}{G^2 m \rho_{c} \ln{\Lambda}}
    \label{t_char}
\end{equation}
as a characteristic time for the evolution of a halo consisting of both PBHs and DM. Here $\rho_c$ is the central density of halo and $\sigma$ is calculated using the Jeans formula \citep{2008gady.book.....B}
\begin{equation}
    \sigma^2 = \frac{1}{\rho_c}  \int_0^{\infty} \rho_H(r') \frac{G M(r')}{r'^2} \, dr',
\end{equation}
where $\rho_H = \rho_{BH} + \rho_{DM}$ is the DM halo density profile. 

For further analysis, it is necessary to determine the halo core collapse time for different fractions of PBHs $f_{PBH}$ in the DM composition. For this, the Fokker-Planck kinetic equation is used, the procedure for numerical solution of which is described in \citet{1980ApJ...242..765C, 2017ApJ...848...10V, Stasenko:2021wej}. In this work for definiteness, the Burkert profile is used as the initial density distribution \citep{1995ApJ...447L..25B} 
\begin{equation} \label{burkert}
    \rho_H = \frac{\rho_c}{\left (1 + r / r_0 \right) \left (1 + r^2 / r_0^2 \right)},
\end{equation}
where $r_0$ is the radius that determines the size of the halo core. We also assume that both PBHs and DM particles are equally distributed in space. That is, it is assumed that the fraction of PBHs matches $f_{PBH}$. However, during the process of halo formation in the central region, the PBH number density can be higher, but this requires careful numerical simulation. In this case, the gravitational waves constraints will be relaxed slightly stronger.

\begin{figure}
	\begin{center}
\includegraphics[angle=0,width=0.5\textwidth]{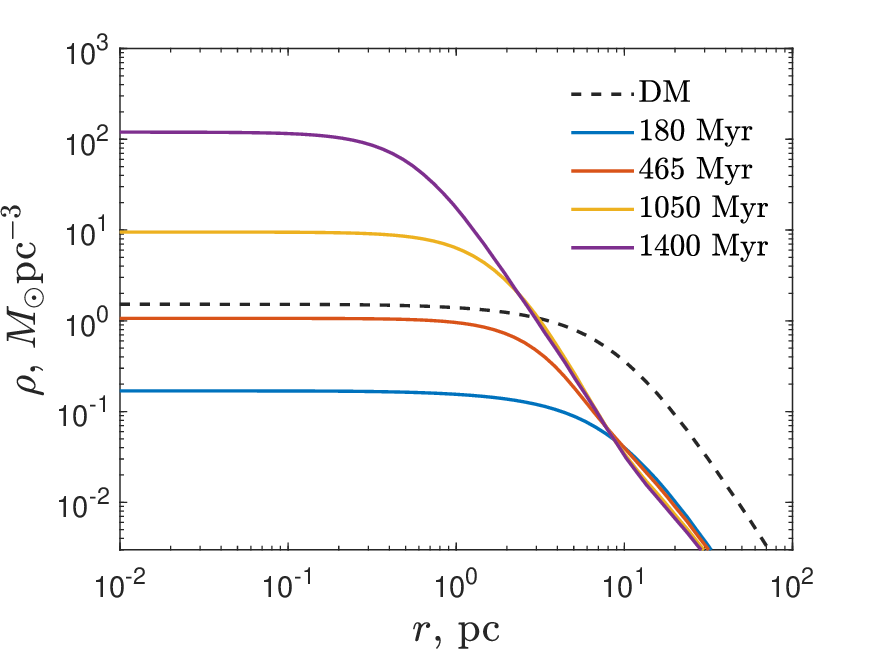}
	\end{center}
\caption{PBH density profile evolution (solid lines) in the halo with mass $M = 3 \cdot 10^4\, M_{\odot}$ formed at $z_f = 20$ and the PBH fraction $f_{PBH} = 0.1$. The dashed line is the density profile of DM particles.}
	\label{rho} 
\end{figure}

As an example, Fig.~\eqref{rho} shows the evolution of PBHs distribution in the halo formed at $z_f = 20$ (it corresponds \mbox{$t_f \approx 180$~Myr).} The PBH fraction was chosen to be $f_{PBH} = 0.1$, and the parameter $r_0$ was chosen so that $R_{vir}/r_0 = 5$ and the halo mass $M = 3 \cdot 10^4 \, M_{\odot}$, which corresponds to a typical halo forming during this epoch, see Fig.~\eqref{gr_Mh}. It can be seen that during the time $t_{cc} \sim 1.2$~Gyr the core of the halo shrinks to $r_c \sim 1$~pc while the density increases to $\rho \sim 100 \, M_{ \odot}$~pc$^{-3}$. We stopped the calculations when the number of PBHs in the central region is $N \approx 20$. Subsequent evolution within the framework of the Fokker-Planck equation would lead to further compression of the core to an infinite density, that is unphysical. In a realistic scenario, as noted above, the core collapse is terminated due to the formation of binaries, which act as a heat source and lead to the expansion of the cluster. However, post-collapse evolution may be more complex due to the dominant role of DM particles composing the dark halo. Nevertheless, it can be noted that the natural formation of PBH clusters is possible, namely, small dark structures with a significant concentration of PBHs in the central regions. 

\begin{figure}
	\begin{center}
\includegraphics[angle=0,width=0.5\textwidth]{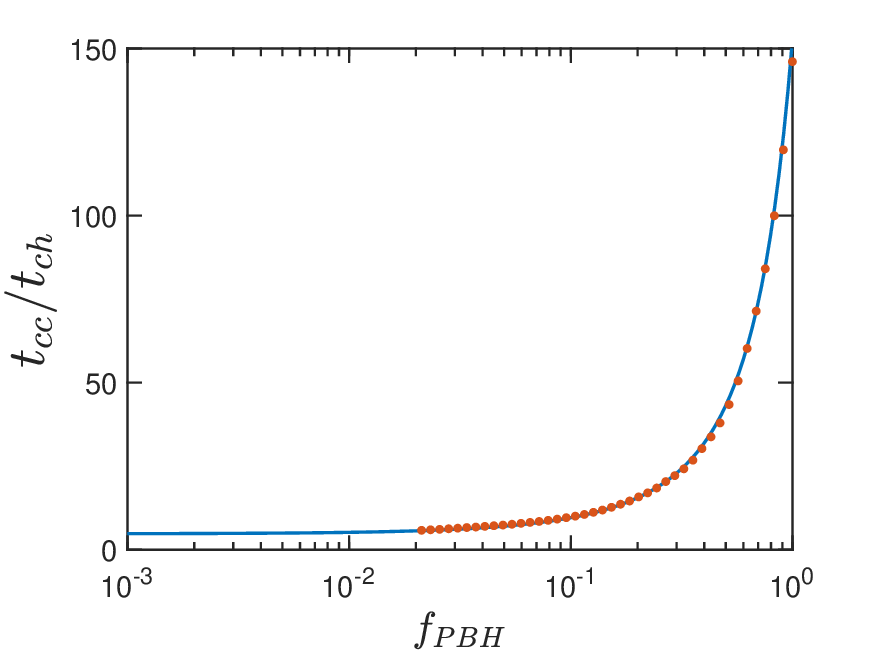}
	\end{center}
\caption{Core collapse time of the characteristic halo mass formed at $z_f = 10$. The red dots are given by the direct numerical solution of the Fokker-Planck equation. The solid curve is given by Eq. \eqref{t_cc}.}
	\label{gr_tcc} 
\end{figure}

Fig.~\eqref{gr_tcc} shows the halo core collapse time in units of characteristic times Eq.~\eqref{t_char} formed at $z_f = 10$ for different PBH fractions $f_{PBH}$. The red dots in the graph are obtained as the result of the numerical solution of the Fokker-Planck equation and are well approximated by the following formula represented by a solid line on the graph
\begin{equation} \label{t_cc}
    \frac{t_{cc}}{t_{ch}} = 15.9 \, \Big (1.3 \, e^{2.1 f_{PBH}} - 1\Big).
\end{equation}

Due to computational difficulties, we considered only halos containing $N_{PBH} \geq 30$, so the Fokker-Planck equation was not solved in the region $f_{PBH} \lesssim 10^{-2}$. In fact, for such fractions $f_{PBH}$, the number of PBHs in dark halos will be less than 30 in this epoch, see Eq.~\eqref{M_char}.
It was found that the core collapse time, expressed in characteristic times Eq.~\eqref{t_char}, is universal and does not depend on the profile parameter $r_0$ and the moment of halo formation, which will be reflected in Fig.~\eqref{gr_Sf}. However, it may depend on the specific choice of the initial density profile \citep{Quinlan:1996bw}. 

The analysis of \citet{Vaskonen:2019jpv} suggests that the core collapse time $t_{cc}$ does not depend on the PBH fraction $f_{PBH}$ and corresponds to a halo consisting only of PBHs, which in our case is $t_{cc} \approx 150 \, t_{ch}$. However, as can be seen from Fig.~\eqref{gr_tcc} for the PBH fraction $f_{PBH} \lesssim 0.1$, such estimation is strongly overestimated. Halos dominated by dark matter evolve faster: as $f_{PBH}$ decreases, the core collapse time becomes constant $t_{cc} \approx 5 \, t_{ch}$, which corresponds to the fact that the PBH dynamics in the halo is mainly determined by dynamical friction.

Let us estimate the characteristic time Eq.~\eqref{t_char} for halos formed at different $z$. To do this, we set \mbox{$\sigma \sim \sqrt{G M / R_{vir}}$} and assume that the halo central density is $\alpha$ times bigger 
than the average matter density in the Universe \mbox{$\rho_c = \alpha \rho_M(z) $} 
\begin{equation} \label{t_ch_est}
    t_{ch} \sim 0.7 \, \frac{10^4}{\alpha}  \left ( \frac{N}{10^3} \right) \left ( \frac{1+z}{10}\right)^{-3/2} \, \text{Gyr},
\end{equation}
where we set $\ln{\Lambda} = 10$ and $N$~--- the number of PBHs in the halo. The choice of $\alpha \approx 10^4$ corresponds to \mbox{$R_{vir}/r_0 = 5$} in the density profile of Eq.~\eqref{burkert}. It can be seen that at the PBH fraction in DM $f_{PBH} \approx 0.1$, halos formed at $z \approx 10$ experience core collapse within the time $t_{cc} \lesssim t_H$. 

\section{PBH binaries} 

In the early Universe, two PBHs at some moment decouple from the Hubble expansion and form a bound binary system, which will subsequently experience the inspiral motion due to the emission of gravitational waves and eventually merge. The lifetime of such a binary is given by \citet{PhysRev.136.B1224}
\begin{equation} \label{t_gw}
    t_{gw}=\frac{3 \, c^5 a^4 j^7}{170 \, G^3m^3},
\end{equation}
where $m$~--- mass of each PBH, $a$ and $j = \sqrt{1-e^2}$ are the semimajor axis and dimensionless angular momentum of the binary system and $e$ is the eccentricity. The distribution of parameters $a$ and $e$ (which is easily converted to $j$) of binaries can be found in \citet{Sasaki:2016jop, Nakamura:1997sm, Eroshenko:2016hmn}.

The merger rate of the unperturbed PBH binaries is given by \cite{Vaskonen:2019jpv, Raidal:2018bbj}

\begin{equation} \label{mr}
    \mathcal{R}_0 = \frac{3.1 \times 10^6}{\text{Gpc}^{3} \, \text{yr}} S \,  f_{PBH}^{53/37}  \left ( \frac{m}{M_{\odot}} \right)^{-32/37},
\end{equation}
where $S$ is the suppression factor arising due to PBH interactions during the formation of a binary
\begin{equation}
    S \approx 0.24 \left ( 1 + \frac{2.3 \, \sigma_M^2}{f^2_{PBH}} \right)^{-21/74},
\end{equation}
where $\sigma_M \approx 0.005$ is the variance of matter fluctuations at the moment of PBH pair formation. 

As shown in the previous section, early DM halos evolve on timescales smaller than the age of the Universe. Therefore, it can be expected that the parameters of the binaries will be perturbed as a result of scattering with other PBHs. Moreover, these PBH binaries formed in the early Universe are highly eccentric $j \ll 1$. The scattering of such binaries with a single PBH will lead to a decrease in eccentricity \citep{Jedamzik:2020ypm}, which ultimately leads to a significant increase of the lifetime. Taking this into account, equation of the merger rate \eqref{mr} will be modified 
\begin{equation} \label{mr_Sf}
    \mathcal{R} = \mathcal{R}_0 S_f,
\end{equation}
where $S_f \lesssim 1$ is the suppression factor that shows the fraction of binaries remaining unperturbed and is calculated in the next section. Scatterings of a binary system with other PBHs can be divided into the following: encounters with large impact parameters, which are of a tidal nature; and close hard scatterings. 

Let us estimate the first. The change in energy in a binary system as a result of such tidal interaction will be \citep{2008gady.book.....B} 
\begin{equation} \label{DE_t}
    \Delta E_{tid} \approx \frac{8 \, G^2 m^3 a^2}{3 \, v^2 b^4},
\end{equation}
where $b$ is the impact parameter and $v$ is the relative velocity (further, for estimates, it is assumed that $v = \sigma$). The rate of energy change is obtained by the standard procedure of multiplying Eq.~\eqref{DE_t} by a factor $n_{BH} \sigma \, 2 \pi b \, db$ and integration over the impact parameter from $b_{min} = a$ to $\infty$
\begin{equation}
    \frac{d E_{tid}}{dt} = \frac{8 \, \pi G^2 \rho_{BH} m^2}{3 \, \sigma}.
\end{equation}
The lower limit of integration was chosen as $a$ based on the consideration that for $b < a$, it is assumed that the scattering is already strong.
It can be shown that an order of magnitude estimate of the change of $j$ is \citep{Ali-Haimoud:2017rtz}
\begin{equation}
    \Delta j \sim \frac{\Delta E_{tid}}{E},
\end{equation}
then, the rate of change in $j$ will be
\begin{equation}
    \frac{d j}{d t} \sim \frac{8 \, \pi G \rho_{BH} a}{3 \, \sigma }.
\end{equation}
Let us estimate the change time of the dimensionless angular momentum by a value of the order of $j$ itself (then the lifetime of binary will increase by the factor $\sim 2^7$, see Eq.~\eqref{t_gw}) as 
\begin{equation}
     t_{j} \sim \frac{j}{ dj / dt} \sim \frac{j \sigma}{ a G \rho_{BH} }.
\end{equation}
At the beginning of the halo evolution, this time is very small (due to the smallness of $\rho_{BH}$), but then the PBH density begins to dominate and the quantity $1 / \sqrt{G \rho_{BH}}$ becomes about the dynamical time in the center of the halo $t_ {dyn} \sim r_{c} / \sigma$, which for $\rho_{BH} = 100$~$M_{\odot}$~pc$^{-3}$ (see Fig. \eqref{rho}) is $t_{dyn} \sim 1$~Myr. Thus, on the time scales of halo evolution, the characteristic time of the change of $j$ can be estimated as 
\begin{equation}
     t_j \sim j t_{dyn} \frac{r_c}{a},
\end{equation}
that is several dynamical times, since $j \sim 10^{-2}$ and \mbox{$a \sim 100$~a.u.} \citep{Ali-Haimoud:2017rtz} and $r_c \sim 1$~pc. Thus, the efficiency of tidal perturbations of binaries regarding changing $j$ becomes important on the timescale of halo evolution.

Another process that may be responsible for the perturbation of the binary parameters is strong scattering with a single PBH. We will assume that the pericenter (the distance of closest approach) for such scattering is $r_p \sim a$. Then, the cross section of such a process will be 
\begin{equation} \label{gr_cs}
    \Sigma \sim \pi a^2 \left ( 1 + \frac{2 \, Gm }{a \sigma^2} \right).
\end{equation}
Since the binding energy of the binaries is greater than the characteristic kinetic energy of the PBHs in the halo, the gravitational focusing approximation is valid (i.e., the second term dominates in Eq.~\eqref{gr_cs}), then the rate of close scattering can be estimated as 
\begin{align}
    \Gamma &\sim n_{BH} \Sigma \sigma \sim \frac{2 \pi a \rho_{BH} G}{ \sigma} \nonumber \\
    &\sim 0.1 \left (\frac{a}{10 \text{au}} \right) \left (\frac{\rho_{BH}}{100 M_{\odot} \text{pc}^{-3}} \right) \left (\frac{1 \text{km}\, \text{s}^{-1}}{\sigma} \right) \, \text{Gyr}^{-1}.
\end{align}

So, it can be seen that strong scatterings are less efficient than long-range tidal interactions, but, nevertheless, they can also make some contribution to the pertu  rbation of binary parameters. Thus, on the timescales of halo core collapse $t_{cc}$, the parameters of the binaries can be significantly perturbed, leading to an increase in the lifetime of the PBH pairs and a reduction in the merger rate.

An important remark should be made. In addition to the merger of binaries formed in the early Universe, direct mergers of PBHs in the dark halo are also possible \citep{Bird:2016dcv, Clesse:2016vqa}. Since halo evolution leads to an increase in the PBHs number density in the central region, direct mergers of PBHs can make a significant contribution to gravitational wave events. However, accurate accounting for these mergers in the modern era requires knowledge of the abundance of such clusters. As an example, it can be seen from Fig.~\eqref{rho} that the maximum PBH number density in the halo is reached by redshift $z \sim 4$. However, in the further process of structures formation, such clusters can be destroyed, which will reduce the rate of direct PBH mergers. We leave this analysis for the next work.

\section{Suppression factor and the PBH merger rate}

To estimate the suppression factor $S_f$ in Eq.~\eqref{mr_Sf}, we use the formalism of \citet{Vaskonen:2019jpv}, the idea of which is to calculate the fraction of PBH binaries that are in halos experiencing core collapse. As shown in the previous section, it is assumed that the lifetime of binaries becomes longer than the Hubble $t_H$ and, therefore, they will not contribute to the merger rate. If at some redshift $z_f$ a halo is formed containing $N_c$ PBHs and experiencing core collapse to 
the redshift $z_c$, then a halo containing a smaller number of PBHs $N < N_c$ (but forming at the same redshift $z_f$) will collapse to redshifts $z > z_c$. The fraction of PBH binaries that will not be perturbed as a result of the evolution of halos formed at $z_f$ is estimated as 
\begin{align} \label{Sf}
    S_f(z) = &1 - \sum_{N = 3}^{N_c} p_N(z_f) \nonumber \\
    & - \sum_{N' > N_c} \left ( \sum_{N = 3}^{N_c} p'_N(z_f) \right) p_N'(z_f).
\end{align}
Here $p_N$ is the distribution function of the number of PBHs $N$ in the halo 
\begin{equation}
    p_N(z) \propto \frac{1}{\sqrt{N}} e^{-N/N^*(z)},
\end{equation}
which follows from the Press-Schechter mass function and it is valid at high redshifts \citep{Hutsi:2019hlw} and $N^*(z) = M^*(z)/m$ is the characteristic number of PBHs in the halo. The physical meaning of the terms in Eq. \eqref{Sf} is as follows: the second term is the probability that the binary is in a halo containing $N$ PBHs, the third term is the probability that the binary is in substructures of the halo with the number of PBHs $N' > N_c$ (but inside this large halo there will be substructures with $N < N_c$). As noted earlier, binaries which are in halos collapsing to the redshift $z$ will not contribute to the merger rate. Therefore, the fraction of these binaries should be subtracted, which is reflected in Eq.~$\eqref{Sf}$.

\begin{figure}
	\begin{center}
\includegraphics[angle=0,width=0.5\textwidth]{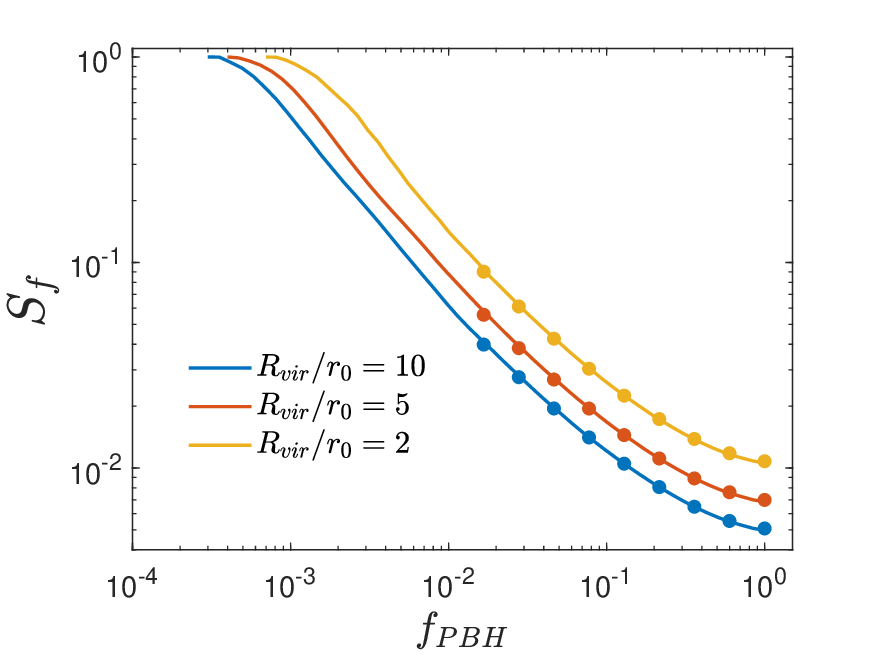}
	\end{center}
\caption{The suppression factor at $z = 0$ calculated for different values $R_{vir}/r_0$: curves from bottom to top correspond to a decrease of the $r_0$.
The solid lines and dots are obtained, respectively, by estimating the lifetime from Eq.~\eqref{t_cc} and by calculating directly the core collapse time using the Fokker-Planck equation.}
	\label{gr_Sf} 
\end{figure}

The further idea is to find the halo formation redshift $z_f$ at which the suppression factor will be minimal to the redshift $z$. This is done as follows: some redshift is taken at which halos are formed. Then, the critical value of PBHs $N_c$ in the halo, which will experience the core collapse to the redshift $z$, is found. Next, we calculate the suppression factor using Eq.~\eqref{Sf}, then we shift along axis $z_f$ in the direction of decreasing $S_f$ until its minimum value is reached. To implement this algorithm, it is necessary to calculate the halo core collapse time. This can be done either directly by solving the Fokker-Planck equation or using Eq.~\eqref{t_cc}. Since we are interested in the modern merger rate, it is further assumed that $z = 0$. 

Figure \eqref{gr_Sf} shows the modern suppression factor calculated for the Burkert density profile \eqref{burkert} for different magnitudes of the parameter $R_{vir} / r_0$. The thick dots correspond to the numerical solution of the Fokker-Planck equation, while the solid curves were obtained using the expression for the core collapse time in Eq.~\eqref{t_cc}. As expected, with the increase of the halo density (a decrease of the parameter $r_0$), the suppression factor $S_f$ decreases, which is due to the fact that the evolution of the halo proceeds more rapidly, see Eq.~\eqref{t_ch_est}. It should be repeated that we do not take into account the influence of hierarchical structure formation, which can lead to tidal disruption of the halo. However, in such process, the central part of the halo will probably continue to evolve, and, in addition, acceleration of the core collapse is possible \citep{PhysRevD.101.063009, Quinlan:1996bw}. 

It is also important to note that the suppression factor $S_f$ does not depend on the PBH mass $m$. This is due to the fact that the characteristic time $t_{ch}$ depends only on the number of PBHs, see Eq.~\eqref{t_ch_est}. On the other hand, the characteristic number of PBHs in the halo $N^*$ also does not depend on the mass of PBH, as can be seen from Eq.~\eqref{M_char}, and also from the fact that Poisson fluctuations are determined only by the number of PBHs $\delta \sim 1 / \sqrt{N}$. 

\begin{figure}
	\begin{center}
\includegraphics[angle=0,width=0.5\textwidth]{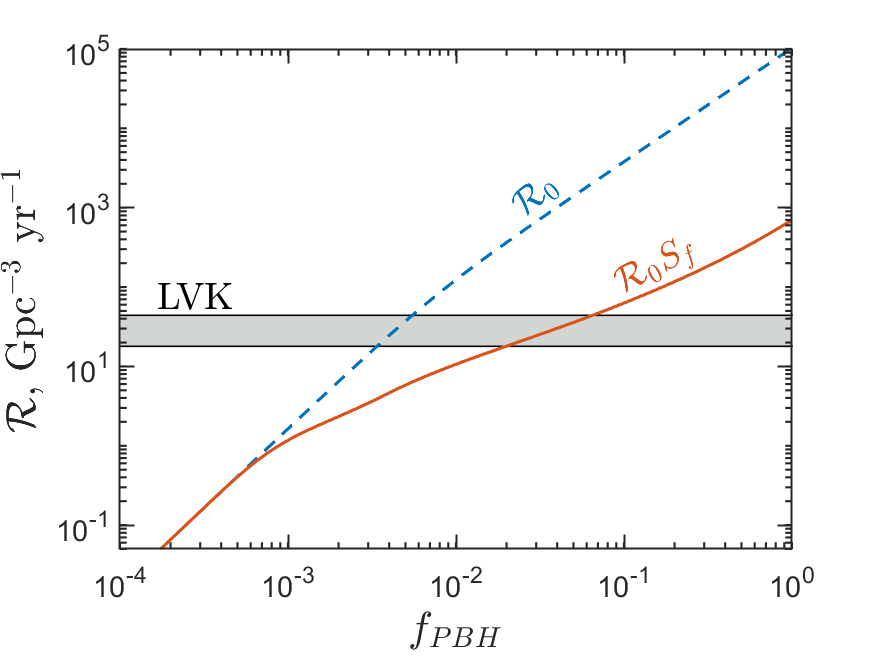}
	\end{center}
\caption{The PBH merger rate with the suppression factor (solid line) and without it (dashed line). The shaded area is the PBH merger rate inferred by the LIGO-Virgo-KAGRA collaboration $ \mathcal{R} = 17.9 \div 44 $~Gpc$^{-3}$~yr$^{-1}$ \citep{KAGRA:2021duu}}
	\label{gr_mr} 
\end{figure}

Fig.~\eqref{gr_mr} shows the PBH merger rate with and without the suppression factor (given by Eq.~\eqref{mr}), where the parameter $r_0$ was chosen as $R_{vir} / r_0 = 5$ for density profile in Eq.~\eqref{burkert}. The shaded area corresponds to the measurements of the LIGO-Virgo-KAGRA collaborations. It can be seen that the fraction of PBHs in DM allowed from the point of view of observations of the BH merger rate is relaxed weakened to $f_{PBH} \sim 0.1$, which corresponds to the constraints on PBHs at masses $\sim 10 \, M_{\odot}$ in the modern Universe on dwarf galaxies \citep{Brandt:2016aco, Koushiappas:2017chw} and lensing \citep{MACHO:2000qbb, Oguri:2017ock}.

In the case of PBH clustering stronger than the Poisson noise predicted in works~\citet{Rubin:2001yw, Khlopov:2004sc} (see also the review of \citet{Belotsky:2018wph}), the constraints are likely to be weakened more. Since in this case, the PBH structures will be formed in a younger Universe. These clusters will experience core collapse on smaller timescales \citep{Stasenko:2021wej}, as a result of which most of the binaries will be perturbed compared to pure Poisson clustering. However, in this case, PBH binaries will already be formed in these clusters through dynamical channels: due to the emission of gravitational waves during close approaches \citep{Bird:2016dcv, Clesse:2016vqa, Stasenko:2021vmm, Garcia-Bellido:2021jlq} and as a result of three-body interactions \citep{Franciolini:2022ewd}. The merger rate of these binaries may dominate over early ones, which requires a separate analysis. 

\section{Conclusion} 

PBHs with the masses $\sim 10 \, M_{\odot}$ are the subject of active discussion regarding strongest constraints on their contribution to the composition of DM ($f_{PBH} \lesssim 10^{-3}$)  due to the observation of gravitational wave signals. This constraint assumes that binaries forming in the early Universe are not perturbed and/or destroyed with time. However, the Poisson initial space distribution of PBHs leads to the early formation of a dark halos. In such halos, the probability of perturbing and/or 
destroying a pair of PBHs is significant. In this work, the dynamics of PBHs in early dark matter structures and its influence on the merger rate were considered. 

To study the dynamics of PBHs in early dark halos, the Fokker-Planck kinetic equation was solved numerically. We have shown that the halo core collapse time essentially depends on the fraction of PBHs in the DM composition $f_{PBH}$. For $f_{PBH} < 1$, the halo evolves much faster than in the case when the entire DM consists of PBHs, which is due to dynamical friction against DM particles. Halo evolution leads to an increase in the density of PBHs in the central region of the halo, which leads to the perturbation of PBH binaries, as a result of which their lifetime can increase significantly.

The result was obtained under the assumption that early halos are not destroyed during the structure formation and their internal dynamics leads to core collapse due to the interaction of PBHs both with each other and with dark matter particles. By calculating the core collapse time, we estimated the suppression factor for the modern merger rate. Namely, it is assumed that if a binary finds itself in a halo that experiences core collapse to the redshift $z = 0$, then it will no longer contribute to the merger rate. This is due to the fact that in the processes of interactions of a binary with other PBHs, its parameters will be significantly perturbed, leading to a binary lifetime will exceed the age of the Universe.

Ultimately, we showed that the constraints on the fraction of PBHs in DM can be relaxed to \mbox{$f_{PBH} \lesssim 0.1$}, which is compatible with the constraints obtained in the modern Universe from dwarf galaxies and lensing. Moreover, in the case of initial PBH clustering, the constraints will probably be weakened more strongly due to the fact that an even larger fraction of binaries will be perturbed.  

\section*{Acknowledgement} 

The authors are grateful for useful discussions by Yu.N. Eroshenko and S.G. Rubin. 
The work was supported by RSF grant \textnumero 23-42-00066, https://rscf.ru/project/23-42-00066/.

\section*{Data availability}
No new data were generated or analysed in support of this research.



\bibliographystyle{mnras}
\bibliography{bib} 





\bsp	
\label{lastpage}
\end{document}